\documentclass[10pt,journal]{IEEEtran}

\usepackage{xr-hyper}
\usepackage{hyperref}
\usepackage{caption}

\DeclareCaptionLabelSeparator*{spaced}{\\[2ex]}
  \usepackage{soul}
  \usepackage{color}
\usepackage[usenames,dvipsnames]{xcolor}

\ifCLASSOPTIONcompsoc
  \usepackage[nocompress]{cite}
\else
  \usepackage{cite}
\fi
\ifCLASSINFOpdf
 \usepackage[pdftex]{graphicx}
\else
  \usepackage[dvips]{graphicx}
\fi
\usepackage{amsmath}
\begin{document}
\title{Scalable End-to-End RF Classification: \\ A Case Study on Undersized Dataset Regularization by Convolutional-MST}

%K. Youssef1, Y. Cai2, G. Schuette1, D. Zhang2, Y. Huang2, Y. Rahmat-Samii2, L.-S. Bouchard1

\author{Khalid~Youssef,~\IEEEmembership{Member,~IEEE,}%Please spell as ``Khalid'' and not ``Khaled'' to avoid citation problems
 Greg~Schuette,
Yubin~Cai,~\IEEEmembership{Member,~IEEE,}
Daisong~Zhang,~\IEEEmembership{Member,~IEEE,}
Yikun~Huang,~\IEEEmembership{Member,~IEEE,}
Yahya~Rahmat-Samii,~\IEEEmembership{Life~Fellow,~IEEE,}
and~Louis-S.~Bouchard
\IEEEcompsocitemizethanks{\IEEEcompsocthanksitem K. Youssef, G. Schuette and L.-S. Bouchard are with the Department of Chemistry and Biochemistry, University of California, Los Angeles, 607 Charles E Young Drive East, Los Angeles, California 90095-1569.\protect\\
E-mail: Kyoussef@ucla.edu, bouchard@chem.ucla.edu
\IEEEcompsocthanksitem Y. Cai, D. Zhang, Y. Huang and Y. Rahmat-Samii are with the Department of Electrical and Computer Engineering, University of California, Los Angeles, 420 Westwood Plaza, Los Angeles, California 90095-1594.\protect\\
Email: rahmat@ee.ucla.edu}% <-this % stops an unwanted space
}

\IEEEtitleabstractindextext{%
\begin{abstract}
Unlike areas such as computer vision and speech recognition where convolutional and recurrent neural networks-based approaches have proven effective to the nature of the respective areas of application, deep learning (DL) still lacks a general approach suitable for the unique nature and challenges of RF systems such as radar, signals intelligence, electronic warfare, and communications. Existing approaches face problems in robustness, consistency, efficiency, repeatability and scalability.  One of the main challenges in RF sensing such as radar target identification is the difficulty and cost of obtaining data. Hundreds to thousands of samples per class are typically used when training for classifying signals into 2 to 12 classes with reported accuracy ranging from 87\% to 99\%, where accuracy generally decreases with more classes added. In this paper, we present a new DL approach based on multistage training and demonstrate it on RF sensing signal classification. We consistently achieve over 99\% accuracy for up to 17 diverse classes using only 11 samples per class for training, yielding up to 35\% improvement in accuracy over standard DL approaches.  
\vspace{0.5in}
\end{abstract}
%\begin{IEEEkeywords}
%Computer Society, IEEE, IEEEtran, journal, \LaTeX, paper, template.
%\end{IEEEkeywords}
}

\maketitle

\IEEEdisplaynontitleabstractindextext

% \fi
\IEEEpeerreviewmaketitle

\IEEEraisesectionheading{\section{Introduction}\label{sec:introduction}}


\IEEEPARstart{W}{ith} applications in defense, retail, healthcare and tomography,~\cite{auto,defense,health,topo,tomo} RF-based sensor systems that can detect, locate and identify targets at long distances and under different weather conditions are being developed.  One example is ground-based and aircraft-mounted radars for the correct identification of targets in the battlefield~\cite{radar1}. %While there are specific challenges to the positive identification of enemy targets due to the lack of knowledge about the target or evasive measures employed, there is also an important need to positively identify friendly targets in order to prevent the tragic striking of one's own forces during so-called ``friendly fire incidents''. %After the invention of radar, attributed to Christian H\"ulsmeyer~\cite{radar2,radar3}, radar-based techniques have proven to be the tools of choice for the detection and tracking of targets since World War II. The ability of radar to operate under a wide range of conditions is due to the wavelength of radar signals, which makes them relatively unaffected by atmospheric and weather-induced attenuation. Ground-based and aircraft-mounted radars play a primary role in target identification.
Another example is ground penetrating radar, which has important applications to mineral resources evaluation~\cite{bib:FRANCKE2012}.
Other examples include advanced personnel screening imagers~\cite{ahmed2012fully}, concealed weapon detection~\cite{agurto2007review}, through-the-wall imaging~\cite{sadeghi2019dort} and vehicular access control using RFID~\cite{bib:almanza2006design}.  The signal received, which is a result of electromagnetic scattering~\cite{bib:mishchenko2015measurement}, is difficult to interpret and model using hand-engineered methods, and this is where blackbox-type approaches such as deep learning (DL) can be useful.

In our previous work~\cite{radar4}, we have presented a new multistage training (MST) approach to DL for RF transmitter identification. MST is a highly distributable structure-parallel 
DL approach that comprises multiple stages of neural network ensembles, each consisting of several small networks, 
which allows efficient utilization of Newton based second-order optimization, and provides a highly effective data-driven regularization. It was originally designed for high-fidelity denoising of magnetic resonance images (MRI) with non-additive noise~\cite{denoising,patent}. The multistage approach allows ``very early stopping'' at each individual stage, 
where a target error is assigned as a stopping criterion in the first stage and is gradually decreased at successive stages. By systematically assigning specific stopping criteria to each stage we can control the speed of convergence in the system as a whole to optimize overall performance and generalization, where a minimal error is reached in the final stage without over-fitting. 
Our second-order MST has proven superior to standard gradient descent based first-order convolutional and deep neural networks (DNN), including MST trained using first-order optimization~\cite{radar4} .

Given that MST has proven superior to other DL methods in RF signal classification~\cite{radar4}, we investigated an extension of MST  employing a convolutional front-end as a feature extraction stage, for a fully automated end-to-end implementation. We refer to the new implementation as convolutional multistage training (C-MST), and we refer to the network architecture as convolutional multistage network (C-MSN). We demonstrate our method on the classification of radar-like signals, 
where a linearly polarized electromagnetic wave illuminates an object which in turn creates a scattered field detected by another antenna. In this experiment we consistently obtain over 99\% accuracy for 17 classes using only 11 samples per class for training, 
%outperforming the closest standard DL approach accuracy by up to 35\%, 
as well as consistency, robustness, wall clock time, and scalability.  

The dataset, which contains many acquisitions of scattered electromagnetic waves acquired in the frequency domain using a network analyzer and measuring the S-parameter for all 17 object classes, will be made publicly available.  The environment mimics radar-like detection of objects under ideal conditions, i.e., without any clutter or motion.  Our contribution in this article is three-fold: 1) providing an overview of existing  techniques, 2) extending our method to include a convolutional front-end thereby enabling us to increase the input size while maintaining the excellent generalization properties of MST, and 3) providing a new benchmark dataset to help standardize the comparison between different algorithms for RF classification. Our algorithm is computationally efficient and allows incremental learning where only part of the network needs to be trained when new targets are added. It can be run on modest computers and may be a good candidate for deployment in the field, for real-time, low-shot-number learning.

%I suggest adding a paragraph that briefly describes our acquired dataset and mention that it will be made publicly available, then explicitly sating our contributions. Something along the lines of: "our contribution in this paper is two-fold; extending our MST method to include a convolutional front-end..., and providing a new benchmark dataset to help standardize the comparison between different algorithms for RF classification..."

\section{Non-DL Methods for Target Identification \label{sec:traditional}}

Target recognition algorithms operate on a measured target signature data for comparison with the previously-derived computer representations of the targets to provide an estimate of the target's identity.  The recognition process is limited by noise in the radar measurements, errors in the generation of signature reference data and the use of classifiers, which usually involves design compromises. A conventional algorithm is template matching using cross-correlation analysis~\cite{radar5,radar6,radar7,radar8,radar9,radar10}. The accuracy of template matching can be improved using statistical pattern recognition techniques designed to determine the class or identity of a measured object by means of the features extracted from the measured pattern or signature~\cite{radar11,radar12,radar13,radar13_2,radar14, radar15,radar16,radar17,radar18, radar19,radar20,radar21,radar22,radar23,radar24}.  Features can be hand-engineered, including polarization enhancement, resonant-frequency poles, multi-path reflection signatures, target structure-induced modulations, microphone effect, jet-engine modulations or features derived from a transform domain representation of the signal~\cite{radar25}.  Features can also be learned from the data using techniques such as $k_n$-nearest-neighbor estimation or Fisher linear discriminant analysis~\cite{duda2012pattern}.  The use of features for recognition provides advantages in reducing the requirements on the size of reference databases. The extracted features are then compared to a database content to maximize the target recognition performance using rule-based Euclidean distance or Bayesian techniques~\cite{radar26,radar27,radar28, radar29,radar30,radar31,radar32,radar33}.  Distance-based methods invoke the sum of all the distance measures for all features and the minimum distance measure is the best assessment of the target's identity. Unlike end-to-end approaches that would ideally extract optimal features directly from the raw data, manual feature selection can be limited by rough approximations and subjectivity. 
%Artificial neural network (ANN) methods~\cite{radar34} and techniques based on support vector machines (SVM)~\cite{radar35} and Markov~\cite{radar36,radar37} models have been studied to perform feature extraction and classifier functions~\cite{radar38,radar39}.   
%The accuracy of existing state-of-the-art methods~\cite{radar34,radar35,radar36,radar37,radar38,radar39} is far from perfect, with performance rapidly degrading at longer ranges where the SNR is low and the radar cross-section (RCS) is small.

\section{Target Identification by DL \label{sec:deep}}

Thanks to the great success of DL methods in recent years in areas such as computer vision and speech recognition, there has been a growing interest to use DL for RF applications. However, unlike computer vision and speech recognition where standard convolutional and recurrent neural network methods have proven to be very effective to the nature of these applications, DL still lacks a standardized approach suitable to the unique nature and challenges of RF. While an ideal target identification method should be able to operate directly on raw data in an end-to-end fashion, fully connected neural network (FCN) implementations trained using standard algorithms are limited by the number of inputs and the number of training samples they need in order to achieve proper generalization and avoid over-fitting. In general, as the input size increases, the number of required training samples increases~\cite{gr0}. Typically, the number of samples  becomes impractical after the input size exceeds only a few hundred inputs. However, digitized and demodulated RF signals are complex-valued and may be thousands of samples long. Standard regularization techniques such as L$_2$-regularization help remedy the problem, but only to a certain extent~\cite{l2}. For this reason, feature extraction remains a crucial step for a FCN to work, and the shortcomings of feature engineering which can be limited by subjectivity and crude approximations inherently remain. 

The introduction of convolutional neural networks (CNN) delivered a tremendous advantage to overcoming this DL problem by limiting the number of connections from the inputs to the network~\cite{cnn, gr0}. Convolutional neurons scan through the entire input but are connected only to a few inputs at a time. This, in addition to the introduction of DL-specific regularization techniques such as dropout layers~\cite{drop}, revolutionized the field of computer vision. CNNs have the ability to automatically extract features from raw inputs, which is a crucial step towards end-to-end implementations. However, due to the immense computational requirements, CNN training algorithms available today are largely based on first-order gradient descent optimization, which poses limitations on the performance and capabilities of CNNs and DL in general. For example, it is not uncommon for an effective computer vision application to require tens or hundreds of thousands of training samples, though it is possible for humans to distinguish between different objects after only few encounters~\cite{comp1, comp2}. A recent study on using CNNs for RF identification shows that several thousand samples per class are needed to achieve high accuracy~\cite{study}. While customized approaches have been proposed for specific applications, performance is very sensitive to hyperparameters and require considerable expert effort to tune for proper parameter selection and tuning. This makes DL implementations narrowly applicable, mainly limited to the exact application and dataset at hand, which poses significant limitations on repeatability. Furthermore, even with suitable hyperparameters, training with different initial conditions, e.g. initial neuron weights and biases, can cause large variations in performance. A typical approach to address this problem is to train one or more network models several times with different initial conditions and use the average output or vote of a committee of networks as the final result~\cite{comt1, comt2, comt3}, which can help improve performance but is ultimately limited by the success rate of the individual networks.  Second-order Newton based optimization methods offer several advantages in robustness to hyperparameters,  efficiency, accuracy,  convergence speed, and require lower network complexity where they can achieve results that are superior to first-order methods with fewer neurons~\cite{gr1, gr2, gr3 ,gr4 ,gr5 ,gr6 ,gr7}. However, second-order optimization is generally deemed unfeasible for training large networks due to the intractable computational requirements of traditional second-order methods.

Provided a large number of training samples is available, most shortcomings of first-order based DL approaches are usually manageable. Unfortunately, obtaining data in large quantities is not trivial in RF problems and can be a difficult, time-consuming, and expensive process.  Synthetically-generated data is often used to compensate for the lack of real data, but it can only approximate the fine details in the signals which poses a limitation on accuracy and scalability. Generative adversarial networks (GAN) have also been used to generate new training data~\cite{goodfellow2014generative}, but GANs can face similar issues depending on application. Recent work~\cite{gan} studying the use of GANs for synthetic aperture radar showed promise at improving the quality of synthetically-generated data. The study also showed that GANs were difficult to train and were not a perfect substitute for real data~\cite{gan}. Artificial neural networks (ANN)~\cite{radar34}, support vector machines (SVM)~\cite{radar35} and Markov~\cite{radar36,radar37} models have been studied to perform feature extraction and classifier functions~\cite{radar38,radar39}. 
Several attempts have been made to adapt standard DL approaches to radar classification with varying degrees of success~\cite{activ, below, fall, sar1, sar2, soli1, soli2, water}.  Hundreds to thousands of samples per class were used for training in order to classify 2 to 12 classes with reported accuracy ranging from 87\% to 99\% where accuracy generally decreases with more classes. Existing DL approaches face problems in robustness, consistency, efficiency, repeatability and scalability.

\section{Implementation of DL by Convolutional MST\label{sec:methods}} 

This paper presents a new variation of MST with a convolutional front end, and its application to the classification of radar-like signals. 
Given the inherent regularization properties of CNNs for large inputs~\cite{gr0,cnn} and their robustness to translational variance, which has been demonstrated in radar ATR~\cite{malmgren2016training}, we have developed C-MST to further improve the performance and applicability of our original MST method.  The addition of CNNs not only helps with translational invariance, but also helps in situations where training data availability is scarce. Unlike our previous work where only part of the RF signal is used (the onset) by the MST, the convolutional front-end allows using the entire RF signal for a fully automated end-to-end implementation.

%ref:
%D. Malmgren-Hansen, R. Engholm and M. O. Pedersen, "Training Convolutional Neural Networks for Translational Invariance on SAR ATR," Proceedings of EUSAR 2016: 11th European Conference on Synthetic Aperture Radar, 2016, pp. 1-4.

We test the new method regularization performance when trained with an undersized dataset (11 samples per class) for a varying number of classes (8 to 17). We term the new method convolutional multi stage training (C-MST).  
%We have developed C-MST to fulfill the need for high accuracy under conditions of limited training data availability.  
Here we describe the implementation of our method as well as conventional DL methods used for benchmarking purposes. The C-MST method and its implementation are described in Section~\ref{sec:C-MST} below. The C-MST method is validated against other DL methods:  CNN, CNN Committee, FCN and FCN Committee, as described below.  In Section~\ref{sec:radexp} we describe experimental methods for acquiring the radar data.

All networks were implemented in MATLAB\textsuperscript \textregistered ~R2019b using the Deep Learning toolbox (The Math Works, Natick, MA) on a CentOS 7-based server featuring two Xeon processors, each with 10 cores, and 128~GB RAM.  Hyperparameters for all methods were selected to maximize generalization. Architecture details and hyperparameters values are provided in the appendix.
The dataset will be made available for download at \hl{https://to\_be\_determined}

\subsection{Convolutional MST (C-MST)\label{sec:C-MST}}

\begin{figure}[h!]
\centering
\includegraphics[width=.7\columnwidth]{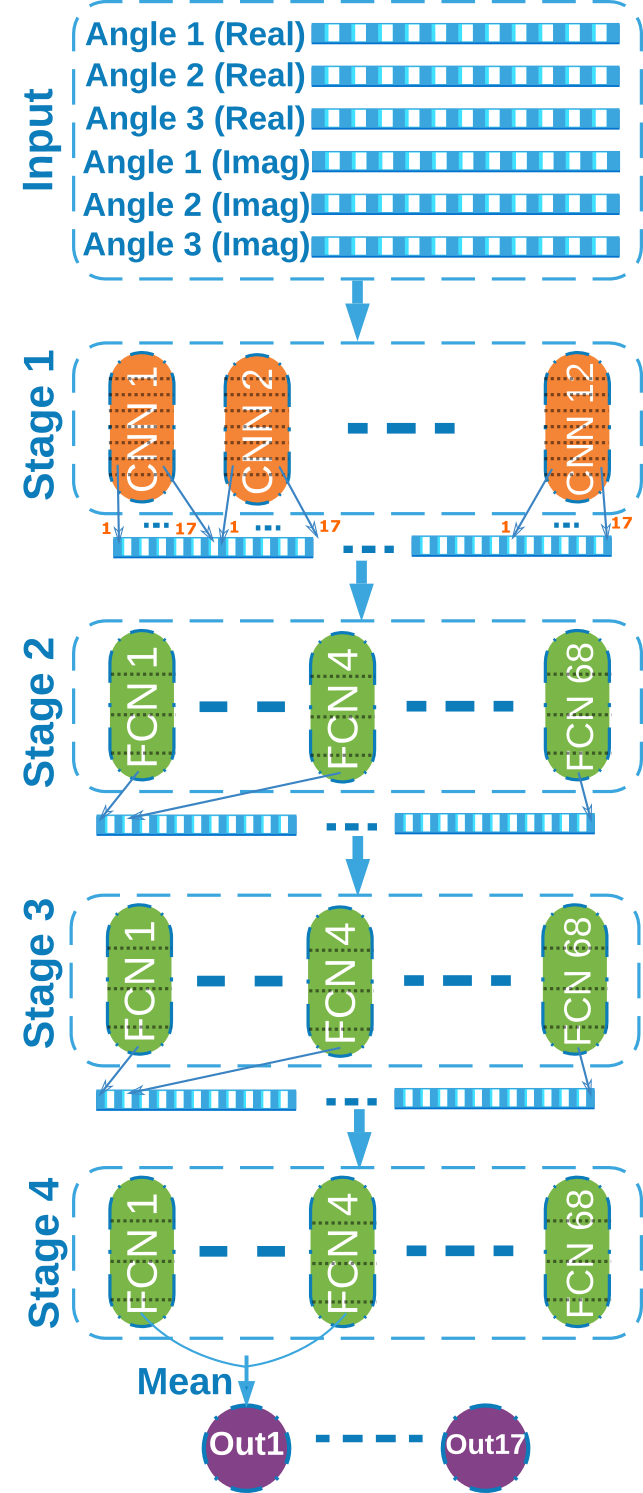}
\caption{\label{fig:4} C-MSN architecture for 17 object classes. A representive vector of the output of each stage in response to one sample is shown for further illustration. Actual stage outputs in response to all samples are shown in Fig. 2}
\end{figure}

\begin{figure}[h!]
\centering
\includegraphics[width=1.0\columnwidth]{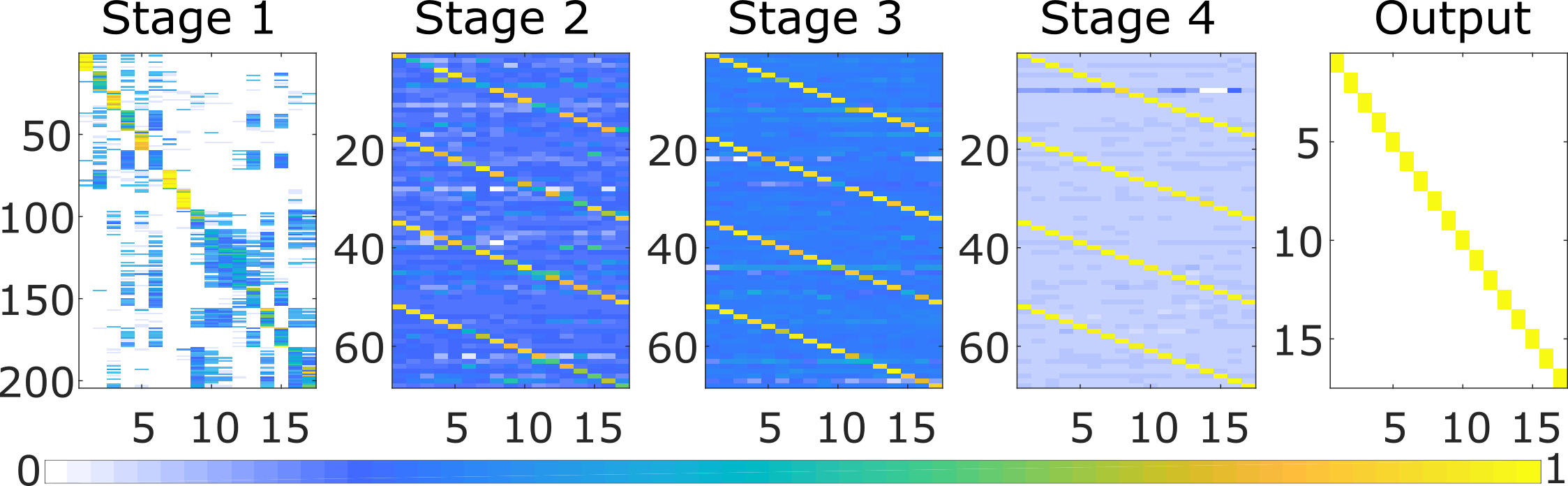}
\caption{\label{fig:5} Outputs from different C-MSN stages. The y-axis is the stacked stage output number and the x-axis is the sample ID. Stage 1 has 12 networks with 17 outputs per network, so the y-axis is from 1 to 204 (17x12). Stages 2 to 4 each have 68 networks with 1 output each. The final output represents a confusion matrix. This figure illustrates the successive evolution of the output, where accuracy gradually improves and noise decreases with each stage.}
\end{figure}

Twelve CNNs constitute the first stage of the C-MSN illustrated in Fig.~\ref{fig:4}. Each CNN consists of 4 inner blocks and one outer block. Each inner block contains a convolutional layer followed by a batch normalization layer, a ReLU layer, and a pooling layer. The outer block consists of a dropout layer, followed by two fully connected layers, a softmax layer, and a classification layer. The architecture details are illustrated in Fig.~\ref{fig:S2} and the hyperparameters for each layer are listed in Fig.~\ref{fig:S-P}.
Each CNN is trained for 3 epochs, and the softmax layer outputs from all CNNs are concatenated to form the input of stage 2. It is important to note that it is the outputs from the softmax layer --- not the output layer --- that are passed on to stage 2. The remaining C-MSN stages are similar to ~\cite{radar4}, where the original MST approach is described in detail. Multiple identical FCNs are stacked together at each stage, where each FCN is randomly assigned different initial conditions. Herein, each FCN consists of two fully connected layers with 10 neurons in each layer. The FCNs in each stage are separately trained for 3 epochs using the second-order Levenberg-Marquardt algorithm; hyperparameters are listed in Fig.~\ref{fig:S-P}. 
Each FCN has one output, and is trained to fire in response to one of the object classes only, where 4 FCNs in each stage are assigned to each object class. In the case of classifying 17 objects, the number of FCNs in each stage is equal to 4 $\times$ 17, which yields 68 FCNs per stage. The concatenated outputs of all FCNs in each stage are passed on as the input to the next stage. In the final stage, the outputs from each group of 4 FCNs corresponding to an object class are averaged together to obtain the final response to that object class. Figure~\ref{fig:5} shows an example that demonstrates the gradual improvement in the outputs of subsequent C-MSN stages.

\subsection{CNN\label{sec:CNN}}

A comparison of C-MST against standard CNN was performed.  A CNN identical to the one described in section~\ref{sec:C-MST} was trained for a maximum of 50 epochs using the Adam optimizer~\cite{kingma2017adam}; Training/validation ratio was set to 70\% and 30\% respectively. Early stoppage based on validation error was used. The weights from the epoch with minimal validation error were selected as the final model weights. Hyperparameters are listed in Fig.~\ref{fig:S-P}. 

\subsection{CNN Committee}

Twelve CNNs identical to the one described in section~\ref{sec:CNN} are trained. Each CNN is assigned different randomly-generated initial conditions. The final output is the majority vote of all CNNs. 

\subsection{FCN\label{sec:FCN}}

The FCN consists of 3 fully connected layers with 200, 200, and 17 neurons, respectively. This is followed by a softmax layer and a classification layer. The FCN was trained for 100 epochs using the Adam optimizer~\cite{kingma2017adam}. Early stoppage based on validation error was used. The weights from the epoch with minimal validation error were selected as the final model weights. Hyperparameters are listed in Fig.~\ref{fig:S-P}. 
L$_2$-regularization was used. The training-to-validation ratio was set to 70\% and 30\%, respectively.

\subsection{FCN Committee}

The FCN committee used twelve FCNs, each identical to the one described in section~\ref{sec:FCN}. Each FCN is assigned different randomly-generated initial conditions. The final output is the majority vote of the outputs from each FCN.

\section{Radar Experiments\label{sec:radexp}}

\subsection{Data Collection}

\begin{figure}[h!]
\centering
\includegraphics[width=1.0\columnwidth]{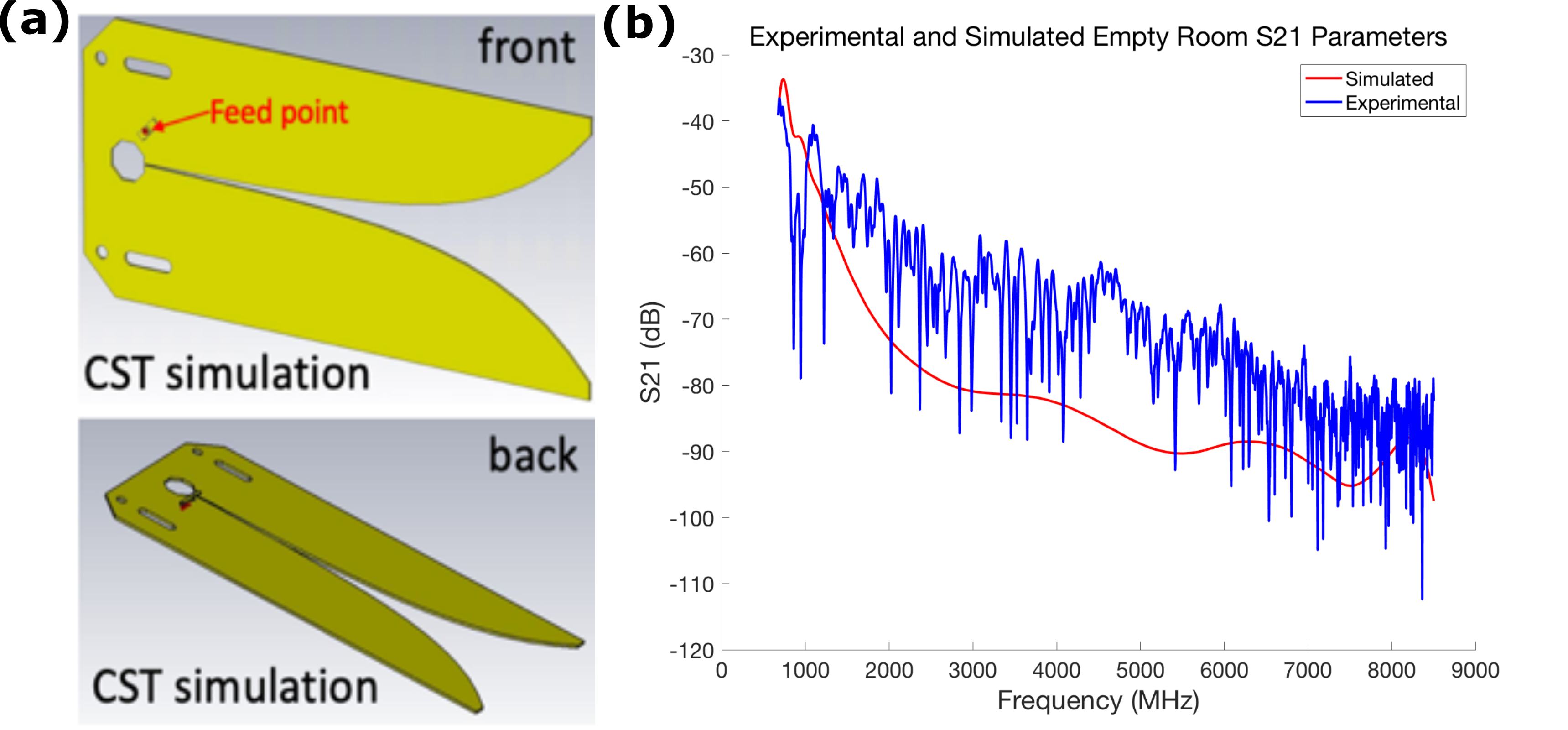}
\caption{\label{fig:SIM} (a) Vivaldi antenna model as used in the CST Studio Suite simulation of the empty chamber. (b)  S21 parameter traces for the empty chamber.  This plot compares simulated vs experimental data. The experimental S21 parameter trace shown was collected on the first day of data collection.}
\end{figure}

\begin{figure}[h!]
\centering
\includegraphics[width=1.0\columnwidth]{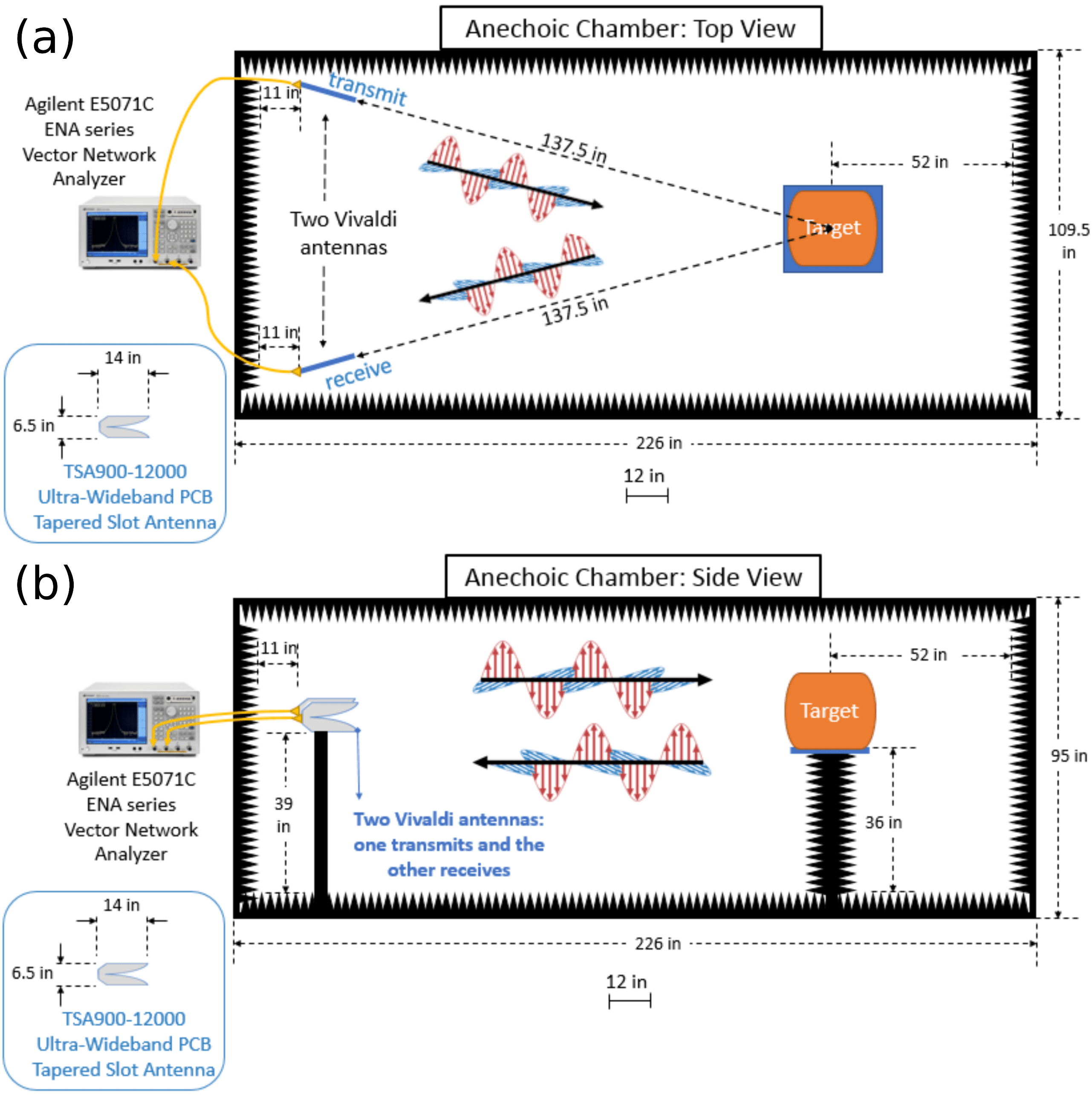}
\caption{\label{fig:1}Schematic of experimental setup.  (a) Top view of the anechoic chamber.  (b) Side view of the chamber.   The ``target'' indicated in the figure represents any of the 17 different objects studied in this project; the 17 objects are shown in Fig.~\ref{fig:S1}. 
Relative proportions are drawn to scale.  Scale bar, 1 ft.}
\end{figure}

Radar data was collected inside an anechoic chamber (band rejection 1 MHz - 10 GHz) using a pair of TSA900 900 MHz - 12 GHz PCB Vivaldi Antennas (RFSPACE Inc., Atlanta, GA) connected to a vector network analyzer (VNA) model E5071C 9 kHz-8.5 GHz ENA Series (Agilent, Santa Clara, CA).  The VNA was operated in S21 parameter mode with ports 1 (Tx) and 2 (Rx) connected to antennas 1 and 2. Frequency sweeping was performed in the range 675 MHz - 8.5 GHz (corresponding to wavelengths in free space: 0.44 m - 3.5 cm) with IF bandwidth 70 kHz, 1,600 points/trace and 512 averages/trace.  Both Vivaldi antennas were mounted vertically to make the readout from the Rx antenna sensitive to waves of the same polarization as the Tx antenna.  The experimental setup is shown in Fig.~\ref{fig:1}.  Each trace (S21 parameter) was saved as a file with 1,600 real/imaginary (complex) data points versus frequency on a linear amplitude scale from the Smith Chart mode of the VNA.  

%The S21 parameter measured in the room without a target (``empty room'') is validated by the electromagnetic simulation tool in CST Studio Suite from Dassault Systemes. The Vivaldi antenna as modeled in CST Studio Suite is shown in Fig.~\ref{fig:SIM}. A method to derive the S21 parameter for bistatic UWB antennas from a simulated S11 parameter~\cite{Yubin_1, Yubin_2} was employed. This approach assumes that the two antennas have identical gain and reflection coefficients. It also assumes the absence of polarization loss.   S21 is calculated from S11 using Eq.~(\ref{eq:2}),

%\begin{equation}
%\vert \text{S21} \vert \equiv \frac{ \vert V_\text{Rx} \vert}{ \vert V_\text{Tx} \vert} = (1 - \vert \text{S11} \vert^2)(\frac{\lambda}{ 4 \pi r})\text{G}_\text{A},
%	\label{eq:2}
%\end{equation}
%where $\lambda$ is the wavelength, $r$ is the distance between receiver (Rx) and transmitter (Tx) antennas, G\textsubscript{A} is the $\lambda$-dependent antenna gain, and $|V_{\text{Rx}}|$ ($|V_{\text{Tx}}|$) is the voltage amplitude at the Rx (Tx) port. 

%In Fig.~\ref{fig:SIM}, we observe agreement between the experimentally-measured vs simulated S21 parameters.  The discrepancy between the two traces is attributed to environmental and instrumentation noises.  

Next, seventeen (17) objects were placed approximately 10 feet from the pair of antennas, which were themselves 5 feet apart.  The triangular configuration was kept fixed and the objects were rotated through 3 angles (0, 45 and 90$^\circ$), and the experiments were repeated 12 times (for each object and angle) on different days over a total period of 3 months.  All targets in subsequent trials and angle rotations were placed at the same locations with intentional positioning errors of up to 10 centimeters in-plane and 5$^\circ$ for the angles.  Other uncertainties in the measurements are due to the limited SNR and possible drifts in the VNA S21 parameter calibration over the 3-month span, as the VNA was calibrated only once on the first day of experiments.  Photos of the 17 objects and their positioning from the antennas' perspective are shown in Fig.~\ref{fig:S1}. 

\subsection{Data Processing}

Since the purpose of this data is to assess end-to-end performance, minimal pre-processing was conducted. Fingerprints used for training the neural networks were created directly from the raw data. Each fingerprint consisted of traces from 3 angles (0, 45 and 90$^\circ$), with each trace consisting of 1,600 real and imaginary values, for a total of 9,600 data points/fingerprint. Real and imaginary components of each trace were normalized separately by subtracting the mean and dividing by the standard deviation. The dataset consisted of 204 fingerprints; 17 object classes with 12 fingerprints per class. Different augmentation and pre-processing  strategies can improve the results for all aproaches. We omit such strategies herein in order to more accurately assess the role of the DL approaches in the performance comparison.

\subsection{Analysis\label{sec:analysis}}

\begin{figure}[h!]
\centering
(a)\includegraphics[width=0.95\columnwidth]{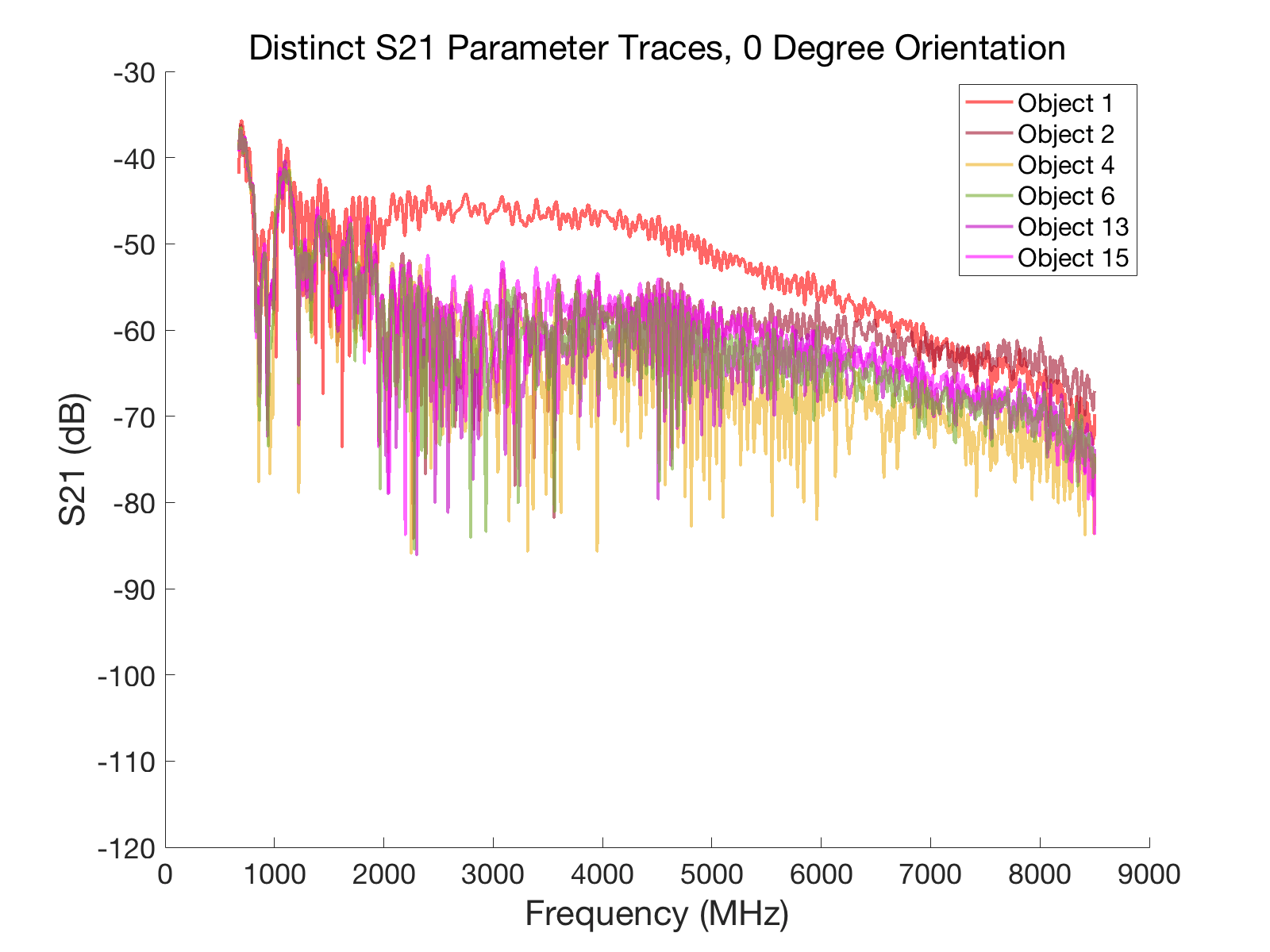}
(b)\includegraphics[width=0.95\columnwidth]{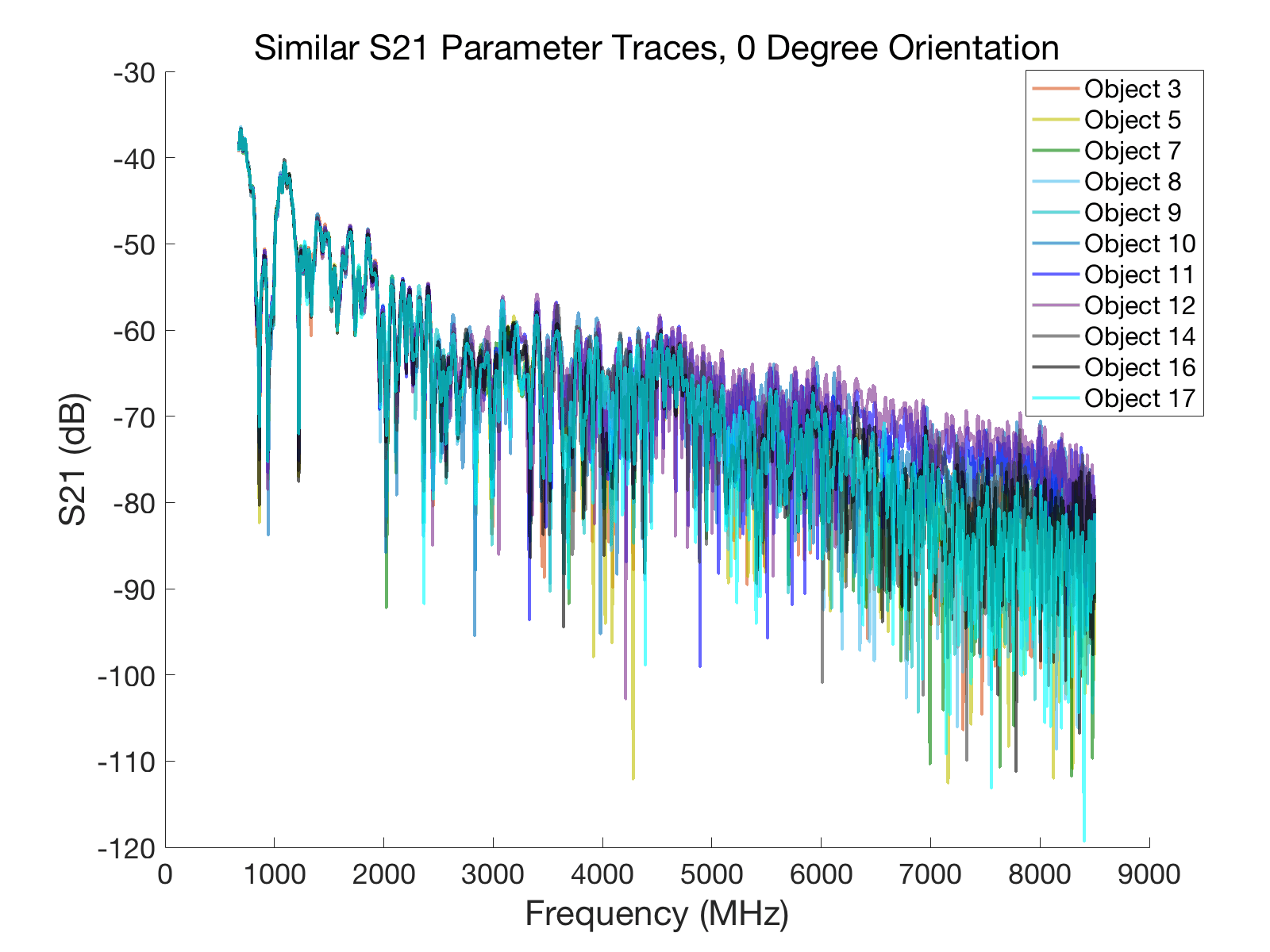} 
\caption{\label{fig:2} S21 parameters (raw data, log-magnitude traces) for all 17 objects at their respective 0$^\circ$ orientation. %(the 0$^\circ$ and 45$^\circ$ orientations are shown in Fig.~\ref{fig:S1}).
Traces show the raw data as acquired without any post processing or averaging.   (a) Several of the targets (objects \#1, 2, 4, 6, 13 and 15) have S21 parameters with obvious differences in the log-magnitude plot.  (b) Other targets (objects \#3, 5, 7, 8, 9, 10, 11, 12, 14, 16 and 17) exhibit S21 parameters with no obvious visual differences.}
\end{figure}

\begin{figure}[h!]
\centering
(a)\includegraphics[width=0.95\columnwidth]{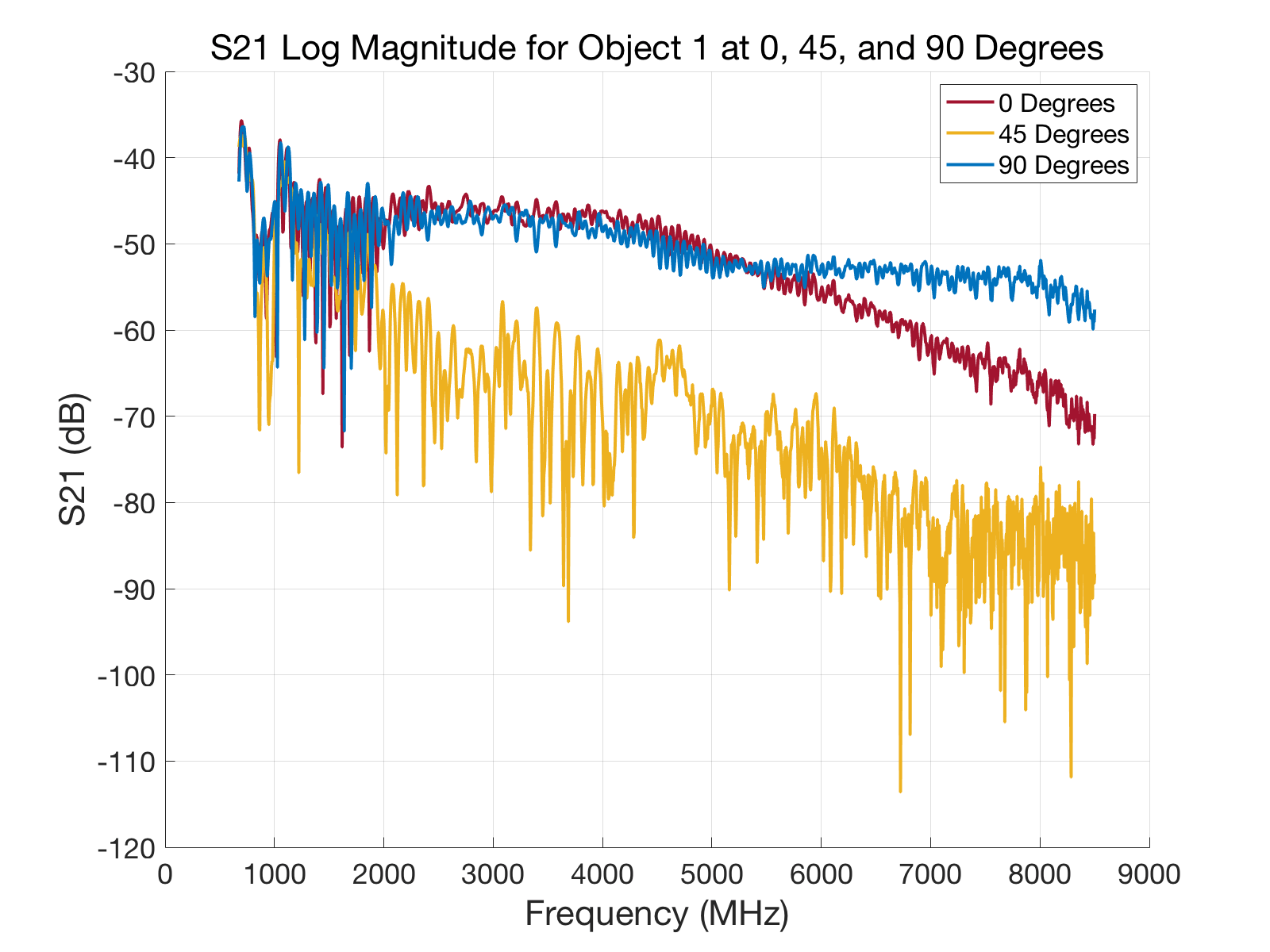}
(b)\includegraphics[width=0.95\columnwidth]{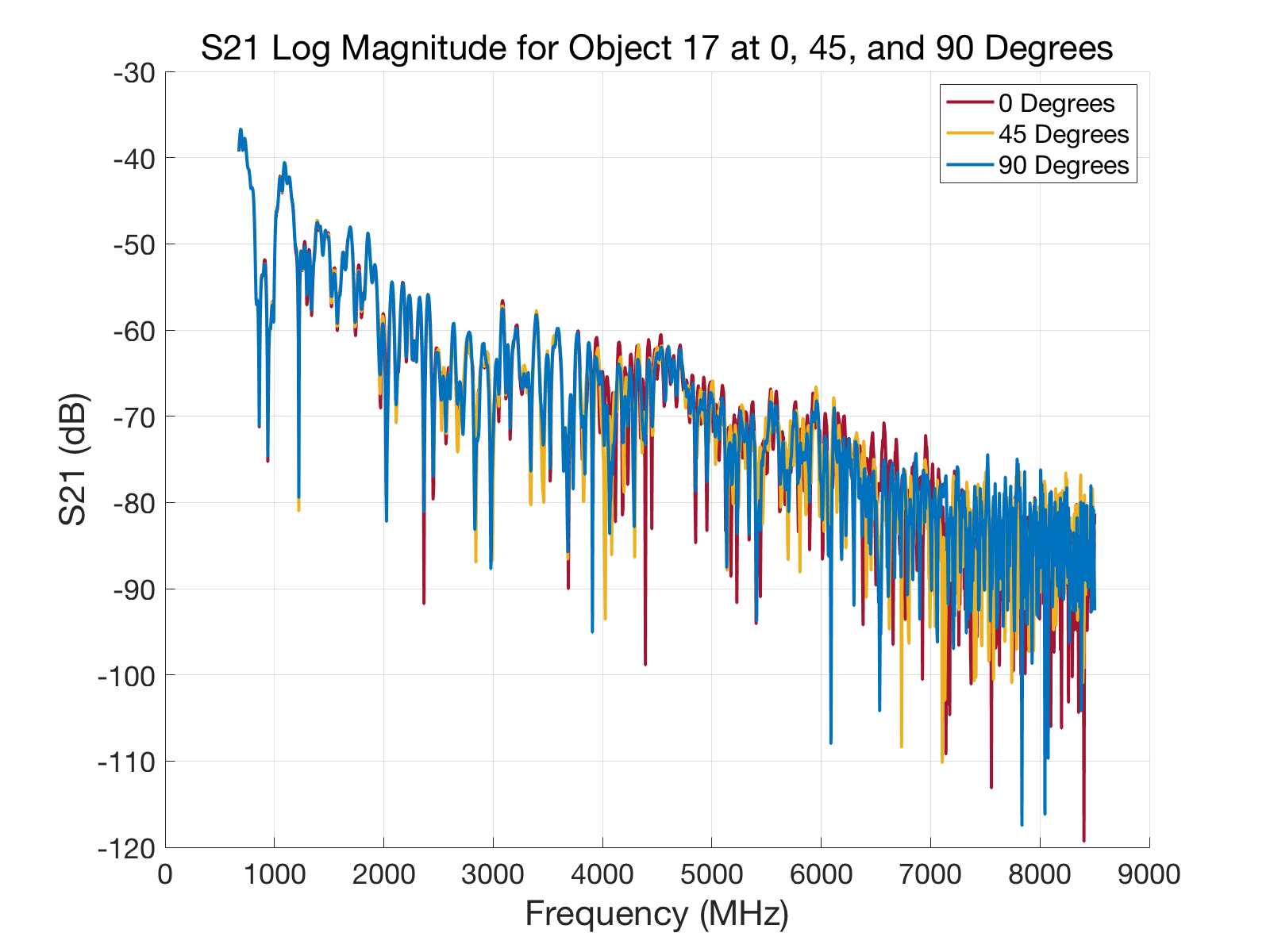} 
\caption{\label{fig:3} S21 parameters for two targets (objects \#1 and \#17) at three different orientations (the objects and a definition of orientation are indicated in Fig.~\ref{fig:S1}). 
No noise removal or averaging was performed on the traces shown. (a) Object \#1 is shown at 0, 45, and 90$^\circ$ orientations, each of which is easily identifiable. Objects \#2, 4, 6, and 13 (not shown here) also demonstrate easily distinguishable orientations. (b) Object \#17 is shown at 0, 45, and 90$^\circ$ orientations, each of which are indistinguishable. Orientations for Objects \#9, 10, 11, 14, and 16 (not shown here) were also visually indistinguishable from their log-magnitude plots. }
\end{figure}

Manual identification of targets from raw data proved difficult, as shown in Fig.~\ref{fig:2}(b) and ~\ref{fig:3}(b).  Figure~\ref{fig:2} shows log-magnitude plots of the S21 parameter for all 17 objects in the 0$^\circ$ orientation. (The objects and our definition of orientation are shown in Fig.~\ref{fig:S1}.)  
In the low-frequency limit of the S21 parameter, most objects were indistinguishable.  Several targets whose S21 parameters are shown in Fig.~\ref{fig:2}(a) displayed traces that can be manually distinguished over the medium-to-high frequency range. Each of these objects were metallic with large cross-sectional areas. This is expected from objects with high conductivity and a large enough area to create conditions for high reflectivity and scattering. On the other hand, several objects, shown in Fig.~\ref{fig:2}(b), have no readily-identifiable characteristics in the S21 parameter that allow us to manually distinguish and identify them.  We note that the high-frequency limit (6-8 GHz) of our experiment was unreliable for manual target identification due to high variance between experiments in cases of low radar cross-sections (RCS).

Figure~\ref{fig:3} examines the angle dependence of the S21 parameter for each object. Certain objects, such as object \#1 in Fig.~\ref{fig:3}(a), reflect the radio waves in different ways depending on orientation. Therefore, once an object is identified, its orientation can be determined from the log-magnitude plot of the S21 parameter. This was the case for relatively large, metallic objects that possess low symmetry and present a unique radar cross-section (RCS) at each orientation. An exception was object \#1, whose 0$^\circ$ and 90$^\circ$ orientations have similar cross sections while exhibiting unique traces (S21 parameters in log-magnitude mode). A possible explanation for this exception is that a crimped copper seam was present on the right and left side of the object for 0$^\circ$ and 90$^\circ$ orientations, respectively. This seam likely reflects and scatters radio waves differently depending on its orientation, allowing each orientation's trace to be manually distinguished. However, other objects displayed little or no clear differences between orientations in the low-frequency, low-noise region, as shown in Fig.~\ref{fig:3}(b). In our experiment, objects with the least-distinguishable orientations tended to be those presenting small RCS areas, regardless of material composition. Small RCSs imply weak scattering intensity and overshadowing of the object's characteristics in the log-magnitude plot. Such objects with traces that cannot be manually identified present a unique challenge to radar target identification.  This is generally the case for targets that present small RCS due to their sizes, reflective properties or range of the measurement.   This apparent difficulty drives the need to develop DL algorithms for radar target identification.  

\section{Results}

\begin{figure}[h!]
(a)\includegraphics[width=0.9\columnwidth]{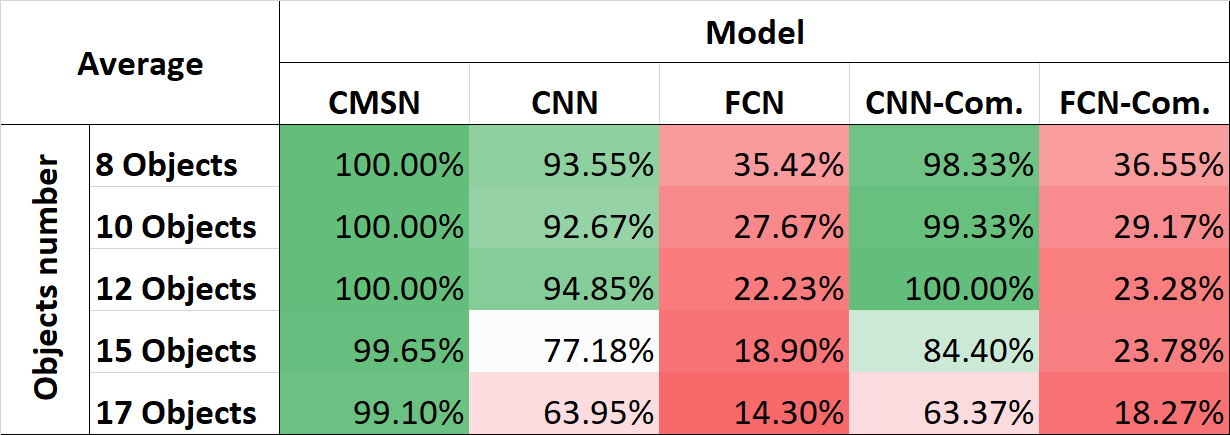} \\
(b)\includegraphics[width=0.9\columnwidth]{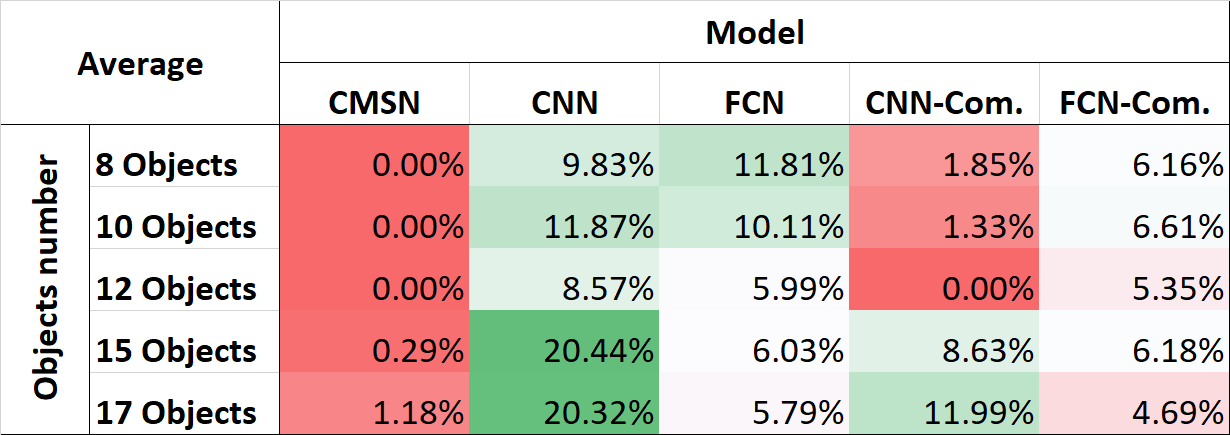} 
\caption{\label{fig:6} Accuracy trends (a) and standard deviation trends (b) for different numbers of object classes are shown. Each point in the graph is calculated from 60 trials. }
\end{figure}

Experiments were designed to test the aspects of accuracy, consistency, robustness and wall clock time of the different DL approaches. As expected, the FCN-based approaches were not suitable for an end-to-end implementation, yet they were included for completeness. The CNN committee was the closest competitor to C-MSN. Therefore, our discussion will focus on comparing CNN to C-MSN results.

\subsection{Accuracy}

\begin{figure}[h!]
\centering
\includegraphics[width=0.6\columnwidth]{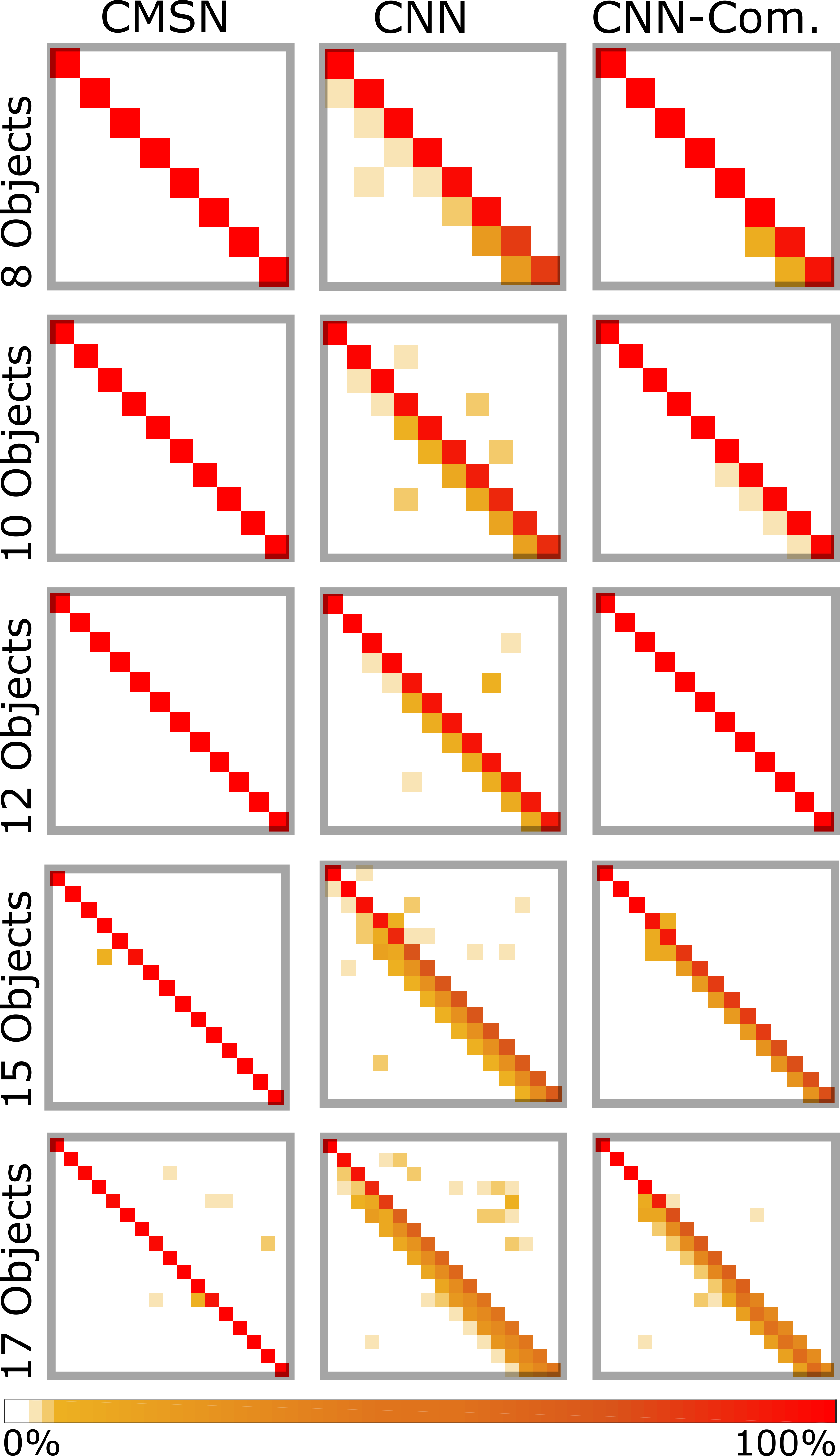}
\caption{\label{fig:7} Confusion Matrices for C-MSN, CNN and CNN committee.  }
\end{figure}

Leave-one-out cross validation (LOOCV) was used for testing the accuracy of each approach. LOCCV is a special case of K-fold cross validation where one sample is left out at a time, and K is equal to the number of samples~\cite{loo}. All combinations of 11 samples per object for training / 1 sample per object for validation were tested. The process was repeated 5 times for a total of 60 trials per accuracy measurement. The average accuracy of the 60 trials was used to calculate each accuracy value shown in Fig.~\ref{fig:6}(a). The confusion matrices for C-MSN, CNN, and CNN Committee are shown in Fig.~\ref{fig:7}.

\subsection{Consistency}

The standard deviation of the individual accuracies from the 60 trials was calculated to assess the consistency of each approach. The results are shown in Fig.~\ref{fig:6}(b). Zero standard deviation indicates that performance remains consistent when training with different initial conditions. With the exception of transfer learning which requires a previous model that has already been trained, initial conditions are typically selected randomly even when following specific initialization techniques. Consistency is a highly-desirable feature that reduces the number of times a model must be trained in order to achieve desirable performance, which increases implementation efficiency. C-MST clearly excels in this aspect, outperforming the other approaches by a large margin. 

\subsection{Robustness to Hyperparameters}

While second-order optimizers are robust to hyperparameter settings, their computational complexity is prohibitive for DL.
Common DL algorithms instead rely on first-order (gradient descent) based optimization, which is sensitive to the settings of hyperparameters such as learning rate and the number of layers and neurons in the network~\cite{xu2020second}. As a distributive training algorithm, C-MST allows the efficient utilization of second-order optimization, making it inherently robust to hyperparameter settings.

Robustness was tested by conducting several classification experiments with varying number of objects. We note, as explained in section~\ref{sec:analysis}, that certain objects are more difficult to classify than others. 
%The hyper parameters of each approach were first optimized for classifying 17 objects. The hyper parameters were then fixed, 
For the purpose of this experiment, the hyperparameters of each approach were first optimized for classifying 17 objects, after which they were fixed
and each approach was tested with a decreasing number of objects, decreasing the level of classification complexity. As seen in Fig.~\ref{fig:6}, C-MSN remains very consistent whereas the performance of both CNN and CNN Committee degrades for less than 12 objects despite the decrease in difficulty.

\subsection{Wall Clock Time}

Despite the seeming complexity of C-MST, which involves hundreds of individual FCNs and CNNs, and although second-order training -- which is typically associated with high computational complexity -- is used, the approach is highly computationally efficient. The computational time in C-MST accelerates virtually linearly with the number of computational cores in a processor, as shown in Fig.~\ref{fig:S3}. 
Given enough cores, C-MST can run in less wall clock time than the CNN and CNN committee approaches. This is due to the highly-distributable nature of C-MST, where individual networks in each stage can be trained independently and in parallel. This is also due to the gradual multistage convergence in C-MST as demonstrated in Fig.~\ref{fig:5}, where individual networks require a very small number of training epochs relative to standard approaches. Furthermore, the partial connectivity in C-MST typically results in a relatively small number of parameters for each individual network in inner stages. %Our experiments were conducted using a dual Xeon processor with 20 CPU cores. (This is mentioned elsewhere)
The average training time for C-MST was 20.2 seconds, while the average training time for CNN was 23.9 seconds, and the average training time for CNN Committee was 31.5 seconds.

\section{Conclusion}
Herein we presented a new DL approach for RF classification, and collected a new benchmarking dataset for proof-of-concept radar applications. The experiments conducted in this study confirm that, while standard CNN approaches can work sufficiently well in some scenarios, their performance drops dramatically as the classification complexity increases. Since classification complexity increases with the number of classes, it is clear that standard DL approaches do not scale effectively in such RF classification applications. In contrast, C-MST is more stable and maintains significantly higher performance across all experiments. Most notably, C-MST maintains a 99\% accuracy in the most difficult classification experiment conducted, representing a 35\% advantage over standard CNN approaches. Additionally, C-MST is robust, computationally efficient and highly distributable. Therefore, C-MST effectively scales with computational complexity and training time as the number of classes increases. We propose C-MST as a scalable end-to-end approach suitable to the nature and challenges of RF data. 

We note that while the main purpose of this study is to compare accuracy and generalization ability of DL approaches when trained with an undersized training dataset, C-MST is equally applicable to pre-processed data and to other radar modalities including 2-D modalities such as synthetic aperture radar, where the input dimensions of the front end CNN stage can be adjusted accordingly. Hyperparameters such as number of stages and number of networks per stage can also be further adjusted according to the problem complexity. 

Furthermore, our newly collected benchmarking dataset will be made publicly available to enable other groups to validate their work independently when applying our method to other challenging radar classification problems. Data collected from pulsed radar in the field generally include additional factors such as polarization, background clutter, time-domain acquisition, variable target-to-radar distance (range), moving targets, additional sources of noise and interference and radar jamming.  The effects of radar clutter were not included here, as the main point of the study was to compare classification accuracy of the new algorithm (C-MSN) to existing, state-of-the-art algorithms under identical conditions.  The numerous advantages offered by our approach will help improve RF-based signal classification performance under these challenging scenarios.

\bibliographystyle{IEEEtran}
\bibliography{refs_v5.bib} 

  \newpage
%\fi



%\bibliographystyle{IEEEtran}

%\begin{thebibliography}{1}

%\bibliographystyle{IEEEtran}
%\bibliography{refs_v3.bib} 

%\bibitem{IEEEhowto:kopka}H.~Kopka and P.~W. Daly, \emph{A Guide to \LaTeX}, 3rd~ed.\hskip 1em plus  0.5em minus 0.4em\relax Harlow, England: Addison-Wesley, 1999.

%\end{thebibliography}

\appendix

% or


\section*{Photos of Objects and Chamber}

\begin{figure}[h!]
\centering
\includegraphics[width=1\columnwidth]{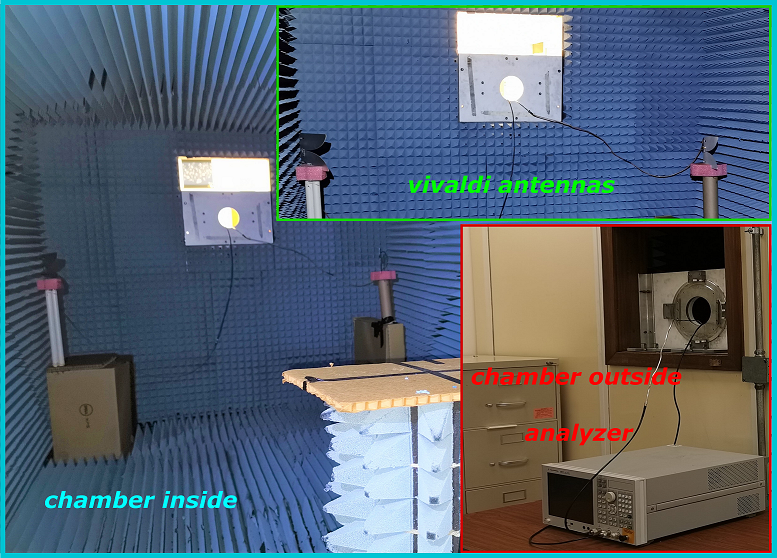}
%\makeatletter
%\makeatletter \renewcommand{\fnum@figure} {\figurename~S\thefigure} 
%\makeatother
\caption{\label{fig:S0} Anechoic chamber and experimental setup. The photo outlined in green shows the two Vivaldi antennas in the lower left and right corners, both oriented vertically, as used during experimental data collection. The photo outlined in red shows the VNA used for all S21 readouts (Agilent model E5071C 9 kHz - 8.5 GHz ENA Series). The photo outlined in blue shows the radar's position relative to the cardboard platform.}
\end{figure}

\begin{figure}[h!]
\centering
\includegraphics[width=.96\columnwidth]{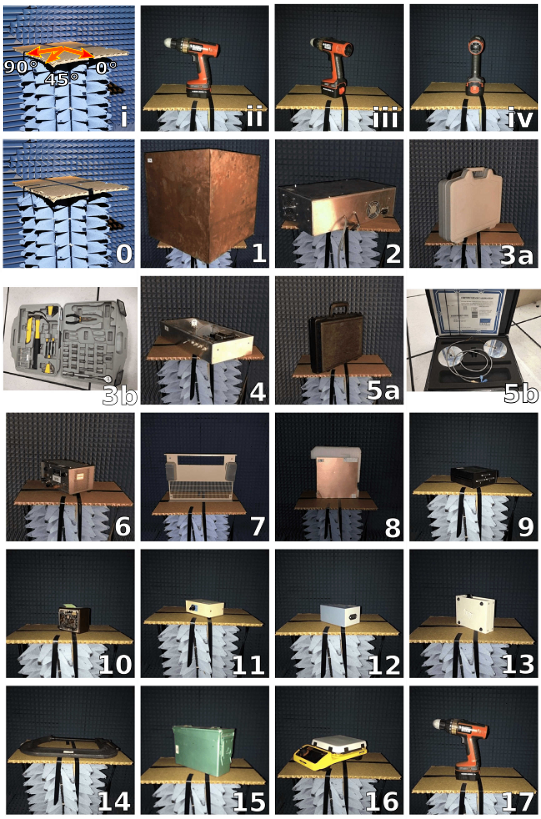}
%\makeatletter
%\makeatletter \renewcommand{\fnum@figure} {\figurename~S\thefigure} 
%\makeatother
\caption{\label{fig:S1} The seventeen targets are shown on the platform. Object size can be gauged by the 47 cm wide and 30.5 cm deep cardboard platform. 0, 45, and 90$^\circ$ orientations are shown in (ii), (iii), and (iv), respectively.  The red arrows (i) depict the orientations on the cardboard. Unless otherwise noted, photos are taken from the radar's perspective at a 45$^\circ$ orientation. The objects vary in size and composition, affecting rRCS. (0) Platform: Roughly 60$^\circ$ from the radar's perspective. (1) Object 1: Empty metal (copper) box with hole on top face. (2) Object 2: Metal box (closed cover) with home-built circuit. (3a) Object 3: Plastic toolbox (closed). (3b) Object 3: Open; data was collected with the toolbox closed. (4) Object 4: Metal box (open cover) with custom circuit. (5a) Object 5: Plastic box (closed). (5b) Object 5: Open; data was collected with the box closed. (6) Object 6: DC power supply (metal cover). (7) Object 7: Front cover of a power amplifier (metal) at 90$^\circ$ orientation. (8) Object 8: Rogers duroid laminate (copper) at 90$^\circ$ orientation. (9) Object 9: Data transfer switch box (with plastic cover). (10) Object 10: Variable capacitor box (metal cover). (11) Object 11: Data transfer switch box (with metal cover). (12) Object 12: Port converter (metallic). (13) Object 13: Data transfer switch box (with plastic and metal cover). (14) Object 14: Vise (metal) at 0$^\circ$ orientation. (15) Object 15: Metal box. (16) Object 16: Chemistry hotplate stirrer. (17) Object 17: Black \& Decker drill at 0$^\circ$ orientation. }
\end{figure}

Seventeen objects were collected from engineering and chemistry laboratories at UCLA for use as radar targets. The objects were selected to present a diversity of sizes, shapes and material composition.  The 47.0 by 30.5 cm cardboard platform seen in the pictures provide a scale for each object's size. All 17 targets were placed in the radar's path and rotated through 3 different angles (0, 45 and 90$^\circ$).  Photos of all 17 rotated objects are shown in Fig.~\ref{fig:S1}, as seen from the perspective of the radar. Photos of the anechoic chamber setup are shown in Fig.~\ref{fig:S0}.

\section*{Signal Strength}

Twelve (12) traces were recorded for each of the seventeen (17) objects, and each of the three (3) orientations per object. Additionally, 112 traces of the empty anechoic chamber were recorded under otherwise equal conditions. Each trace (S21 parameter) was recorded as a string of complex numbers (real, imaginary) representing the complex-valued signal amplitude as a function of sweep frequency.   Signal-to-noise ratios (SNR) and relative radar cross sections (rRCS) were calculated for each of the 17 objects and their three orientations using this data and tabulated below (Table~\ref{tab:1}).   Radar cross sections (RCS) are provided as dimensionless numbers between 0 and 1 (rRCS).  SNR values reflect the maximum amplitude in the frequency domain of the S21 parameter, whereas rRCS values reflect the total signal over the same domain. This is analogous to the pulse radar echo amplitude.  Uncertainties in the values represent sample standard deviations, which were calculated using the 12 traces for each object (and orientation).

From each complex-valued trace for each object (and orientation), the magnitude of the complex data was computed and stored in vectors.  Each trace vector was 1,600 points long, corresponding to linearly spaced frequency values in the range 675 MHz - 8.5 GHz.  For the empty anechoic chamber data, all 112 traces were averaged to provide a clean (low noise) trace.  This low-noise trace was used to subtract the background signal for each object (and orientation) in order to produce traces whose features only reflect the characteristics of an object.

From the background-subtracted traces, SNR was calculated as follows. The location of maximum signal was identified and 100 nearby points were averaged to reduce noise. For each trace, a flat region was identified in order to estimate the noise as the standard deviation of 100 points taken in the flat region. SNR was computed as the above-mentioned signal strength divided by the standard deviation of the flat region.  The resulting value was inserted into Eq.~(\ref{eq:1}), 

\begin{equation}
\left. \text{SNR} \right|_{\text{dB}} = 20 \log_{10} \left( \frac{\text{S21}_{\text{target removed}}}{\text{S21}_{\text{with target}}} \right).
	\label{eq:1}
\end{equation}

\noindent to yield a value in dB.  rRCS values were estimated from the background-subtracted traces by taking the area under each curve, which represents the total radar signal (integrated across all frequencies).  The integral was computed in MATLAB using the trapezoidal rule.  This area represents the ``pulse amplitude'' in a pulsed radar echo experiment.  These pulse amplitudes were then averaged for each object-orientation combination and normalized to the largest object-orientation average to yield a number between 0 and 1 (what we call the ``relative'' RCS). 

\begin{table}
\tiny
\begin{tabular}{lllllll}
Object & SNR, 0$^\circ$ & SNR, 45$^\circ$ (dB) & SNR, 90$^\circ$ (dB) & rRCS, 0$^\circ$ & rRCS, 45$^\circ$ & rRCS, 90$^\circ$ \\ \hline
1 & 20(10) & 13(8) & 19(9) & 0.9(3) & 0.10(2) & 1.0(4) \\
2 & 14(7) & 14(7) & 8(7) & 0.34(5) & 0.10(2) & 0.4(1) \\
3 & 13(7) & 14(8) & 12(7) & 0.10(2) & 0.10(2) & 0.14(2) \\ 
4 & 12(7) & 14(7) & 11(6) & 0.22(6) & 0.09(3) & 0.17(3) \\ 
5 & 14(8) & 13(7) & 14(8) & 0.10(2) & 0.10(2) & 0.17(4) \\ 
6 & 12(8) & 13(7) & 10(6) & 0.23(3) & 0.10(3) & 0.17(2) \\ 
7 & 13(8) & 13(8) & 14(9) & 0.10(3) & 0.10(2) & 0.5(2) \\ 
8 & 12(7) & 14(8) & 13(8) & 0.10(3) & 0.10(3) & 0.20(3) \\ 
9 & 12(6) & 14(7) & 10(6) & 0.10(3) & 0.10(3) & 0.11(3) \\ 
10 & 10(6) & 13(7) & 12(7) & 0.12(3) & 0.09(3) & 0.09(3) \\ 
11 & 10(6) & 14(8) & 11(6) & 0.12(2) & 0.09(3) & 0.10(3) \\ 
12 & 10(6) & 13(7) & 10(6) & 0.14(3) & 0.09(3) & 0.11(3) \\ 
13 & 11(7) & 12(7) & 10(6) & 0.23(2) & 0.10(3) & 0.11(3) \\ 
14 & 11(6) & 13(7) & 12(7) & 0.11(3) & 0.09(3) & 0.10(3) \\ 
15 & 14(8) & 11(6) & 11(7) & 0.44(5) & 0.10(3) & 0.14(2) \\ 
16 & 11(7) & 13(7) & 11(6) & 0.10(3) & 0.10(3) & 0.10(3) \\ 
17 & 12(7) & 13(7) & 13(7) & 0.10(3) & 0.10(3) & 0.10(3)  \\ \hline
\end{tabular}
%\makeatletter
%\makeatletter \renewcommand{\fnum@table} {\tablename~S\thetable} 
%\makeatother
\caption{\label{tab:1} SNR and rRCS values for each object (1--17) and their three orientations. The standard deviation of each signal is shown in parentheses. }
\end{table}

\section*{CNN Architecture and Hyperparameters\label{sec:hyper}}

\begin{figure}[h!]
\centering
\includegraphics[width=.85\columnwidth]{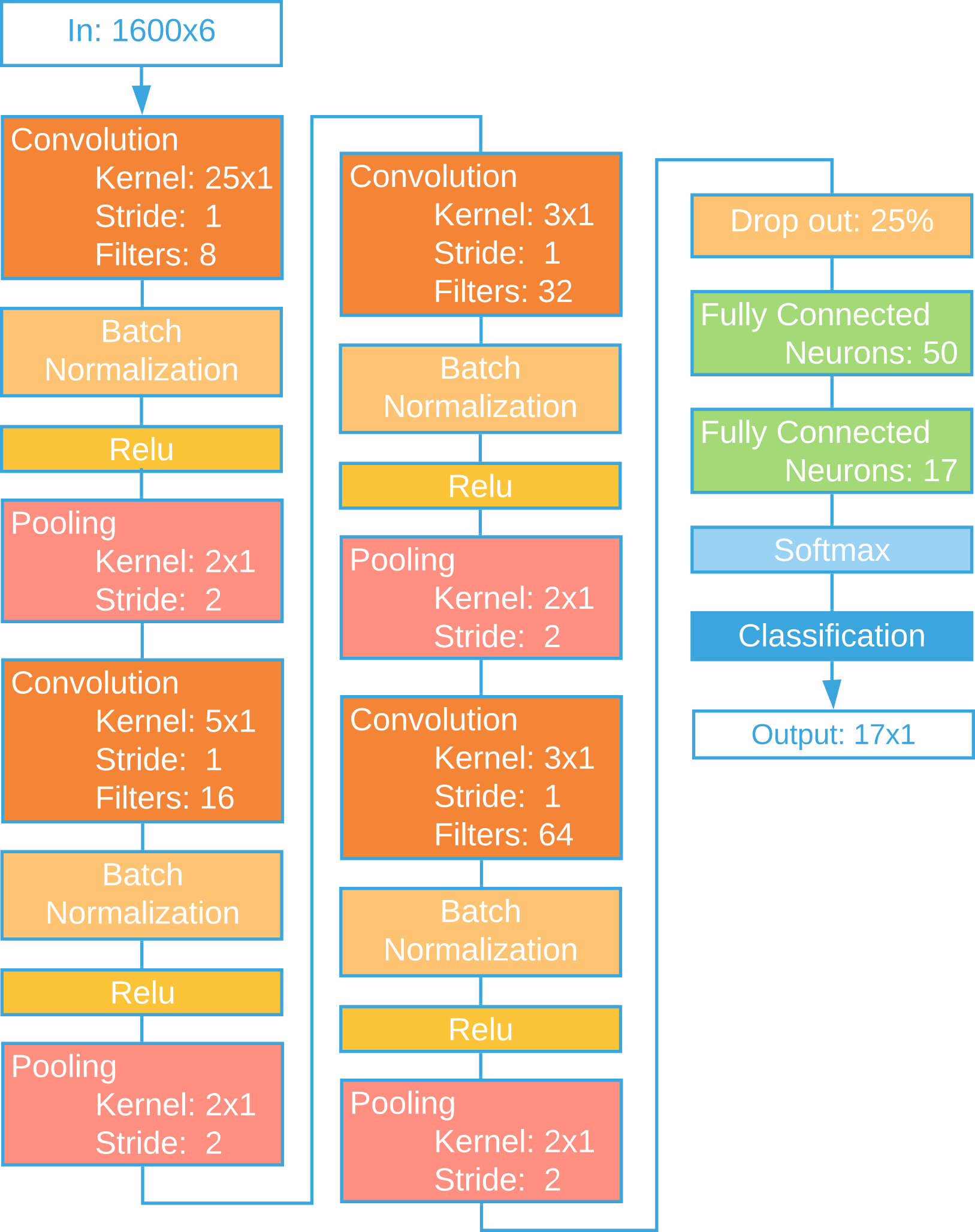}
%\makeatletter
%\makeatletter \renewcommand{\fnum@figure} {\figurename~S\thefigure} 
%\makeatother
\caption{\label{fig:S2} CNN architecture and hyperparameters. }
\end{figure}

\begin{figure}[h!]
\centering
\includegraphics[width=0.4\columnwidth]{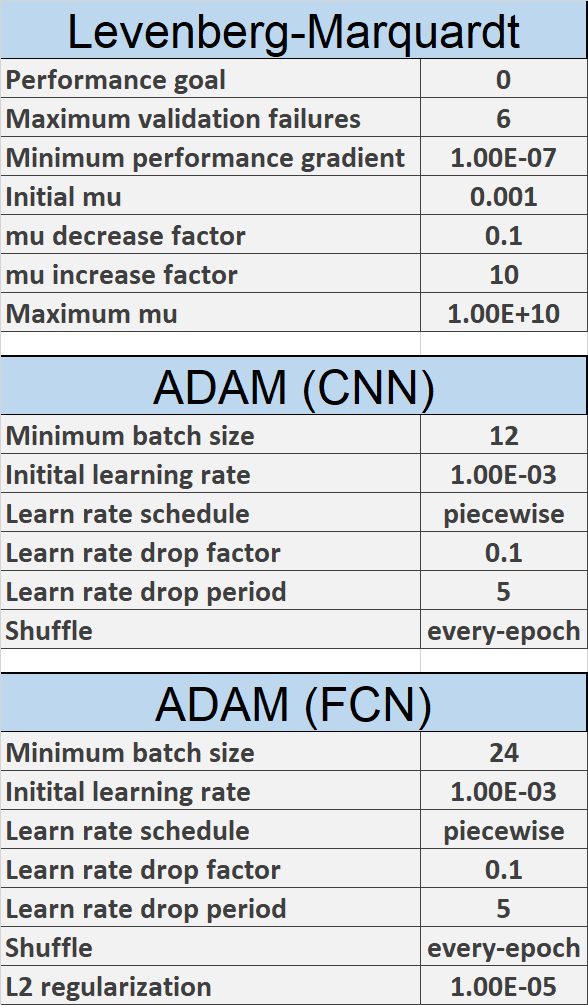}
%\makeatletter
%\makeatletter \renewcommand{\fnum@figure} {\figurename~S\thefigure} 
%\makeatother
\caption{\label{fig:S-P} Levenberg-Marquardt and Adam parameters. }
\end{figure}

The detailed CNN architecture and hyperparameters are shown in Fig.~\ref{fig:S2}. The hyperparameters used for Adam and Levenberg-Marquardt are provided in Fig.~\ref{fig:S-P}.

\section*{Wall clock time}

\begin{figure}[h!]
\centering
\includegraphics[width=.8\columnwidth]{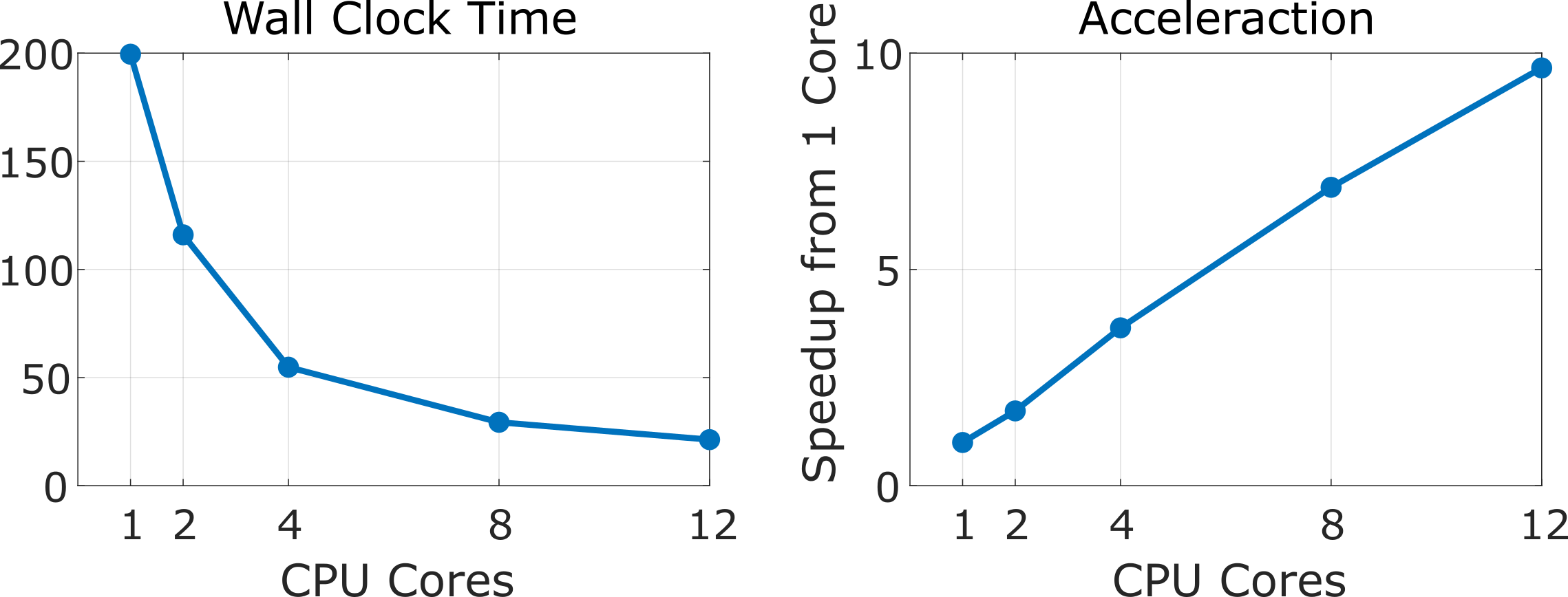}
%\makeatletter
%\makeatletter \renewcommand{\fnum@figure} {\figurename~S\thefigure} 
%\makeatother
\caption{\label{fig:S3} Wall clock time and acceleration. }
\end{figure}

Fig.~\ref{fig:S3} shows wall clock time and speedup relative to 1 CPU core. Measurements were obtained by activating different number of CPU cores and measuring the training time.

\section*{Detailed Results}

\begin{figure*}[h!]
\centering
\includegraphics[width=2.0\columnwidth]{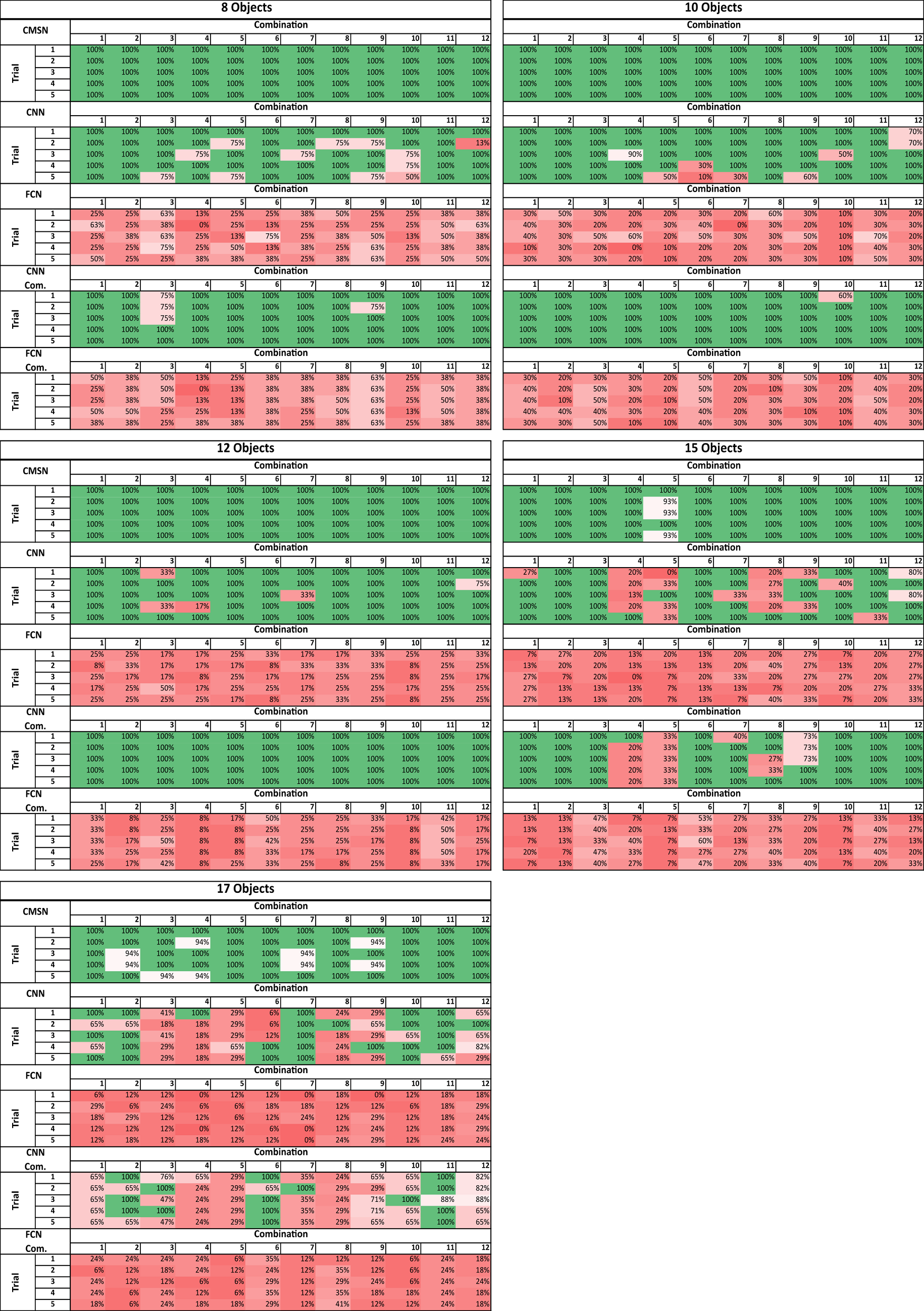}
%\makeatletter
%\makeatletter \renewcommand{\fnum@figure} {\figurename~S\thefigure} 
%\makeatother
\caption{\label{fig:S4} Detailed classification results obtained by different machine learning algorithms. }
\end{figure*}

Full results from each individual iteration are shown in Fig.~\ref{fig:S4}. These are the results used to calculate the values of the average accuracy and standard deviation in Fig.~\ref{fig:6}.

%\vfill


\end{document}